# A NETWORK FOR SOLAR OPTICAL OBSERVATIONS IN BULGARIA

Rositsa Miteva[1], Pencho Markishki[1], Werner Pötzi[2], Momchil Dechev[1], Dragomir Marchev[3], Samuil Petrov[3], Georgi Latev[1], Hristo Kirov[4], Kamen Kozarev[1]

[1]*Institute of Astronomy and National Astronomical Observatory –
Bulgarian Academy of Sciences*
[2]*Institute of Physics, Kanzelhoehe Observatory for Solar and Environmental Research, University of Graz, Austria*
[3]*Astronomical Center "Konstantin Preslavsky" University of Shumen*
[4]*St. George International School and Preschool – Sofia, Bulgaria*
*e-mail: rmiteva@nao-rozhen.org*

*Abstract: The report presents a new initiative for the development of a network of ground-based stations for solar observations in the optical range. Three separate locations in Bulgaria have installed instrumentations for solar dedicated observations. The currently used telescopes (type, mounting, guiding systems) and designated filters (white-light, H-alpha) are described in detail. Test images are also included. Future plans for improvements are briefly discussed.*

## Introduction

Historically, the first telescope observations of the Sun, similarly to the other astronomical objects, were performed in the optical range. The observers are as follows, according to https://chandra.harvard.edu/edu/formal/icecore/The_Historical_Sunspot_Record.pdf:

- The very first solar observer is regarded to be Thomas Harriot (England) based on the notations and solar drawings in his journal dated 1610-12-08.
- In the following year, 1611, reports of sunspots were published in Germany by David and Johannes Fabricius, however no drawings were found.
- In 1612 Galileo Galilei (Italy), with his regular solar observations and drawings (https://galileo.library.rice.edu/sci/observations/sunspot_drawings.html), established the subject as an integral part of astronomy.

Contemporary solar monitoring relies on the data from numerous ground-based stations and spacecraft sensors. One notable ground-based station is Kanzelhöhe Solar Observatory – KSO (Austria, https://www.kso.ac.at/index_en.php) which will be used as an example for the development of the Bulgarian stations. Detailed information about the monitoring program and data products at KSO can be found in Pötzi et al. (2021). Among their data products are white-light and H-alpha (6563 Å) observations which will be also developed at the Bulgarian stations. Regular H-alpha observations are provided (as an additional science output) also by a Global Oscillation Network Group, https://gong.nso.edu/.

In order to avoid the night-time interruptions and weather conditions, one needs to invest in the development, launch and maintenance of space-borne observations. Satellites have been providing nearly uninterrupted streams of images, as long back as 1995 with the launch of the Solar and Heliospheric Observatory (SOHO) spacecraft orbiting around L1. An important recent addition to the solar-dedicated instrument fleet is the Solar Orbiter (launched in 2020) with the aim to reach as close as 0.284 AU and look above the ecliptic. A subset of solar images at various wavelengths are released daily at the well-known database for solar observations https://solarmonitor.org/.

White light (continuum) observations provide information mostly on the sunspots (i.e., areas of magnetic field concentrations emerging through the photosphere which appear dark due to decreased heat transfer, Solanki, 2003). The H-alpha line probes the solar chromosphere, where phenomena such as filaments/prominences, plages, flares, fibrils/spicules (and sunspots) are visible (Pötzi et al. 2021). Both wavelengths are complementary and the respective filters are aimed to be installed in the Bulgarian stations.

The construction of an observatory is a laborious task but should be regarded as an investment in the development of a know-how and human potential. In Bulgaria, we will use the available infrastructure at three separate locations (dome, telescope, internet connection, etc.) and build upon it. It is also advantageous to set up a well-targeted observer program with the option of rapid changes that are project or personnel dependent. Moreover, the development of an optical network will serve as a complementary monitoring program to the observation in different wavelengths, e.g. in the radio-domain. Currently, there are ongoing efforts for the building of a Bulgarian station of dual radio antenna field (https://lofar.bg/bg/), part of the European radio telescope, Low Frequency Array (LOFAR). A well-established optical solar network can be assembled as a trigger for the radio observations. This serves as an additional motivation for the development of various software tools, automatic pipe-lines and machine learning methods (e.g., Ramos et al. 2023). Moreover, the optical observatories can be a locus for pupil, student and young scientists educational and research programs, as well as outreach programs for the general public.

The aim of the presented initiative is to outline the available infrastructure housing the solar-dedicated instrumentation and filters as well as to showcase the first test results.

**Instrumentation and Results**

Below we present a brief description of each station, currently available instrumentation and test images. The diameter of the main mirror/lense (or aperture) is denoted with D and the focal length of the telescope - with f.

The first observing station of the network is the Astronomical Observatory (AO) – Belogradchik, located inside the city (elevation of about 522 m). Typical conditions for light pollution apply. Nevertheless, the observatory is used for regular night-time observations and recently also for solar observations. The telescope, rejection filter, H-alpha filter and first images were already described in Markishki et al. (2025). Here we present an improvement in the filter system, namely the Lunt H-alpha filter (mounted off-axis on a plate, in front of the telescope tube), see Fig. 1 (left). For completeness, the current instrumentation is outlined below:

- Telescope: Celestron C11 – Schmidt-Cassegrain – C11 Optical Tube Assembly, D = 279 mm, f = 2800 mm, focal ratio: f/10
- Mount/guiding: VX AVX GoTo equatorial mount with stable guiding system
- Filter: LUNT LS60F H-alpha (i.e., 60 mm diameter) with blocking filter B600d1
- Camera: Common Vision Spark Series, SP-12401M-USB8, pixel size is 3.45 × 3.45 μm (289.8 px/mm), sensor's active area is 14.19 × 10.38 mm, the maximum frame rate of 23.4 fps with 8-, 10-, or 12-bit monochrome output, and the maximum exposure time during operation is 1/23.4 s (0.042735 s).

A test image was taken on 11-07-2025 and shows a sunspot structure (Fig. 1, right). According to the solar monitoring database https://solarmonitor.org/index.php?date=20250711 this is active region (AR) 14136. However, a strong interference pattern is also present, compromising the overall quality. Due to the system limitation (off-axis location of the filter) there is the option of cropping the image around the center of the sensor.

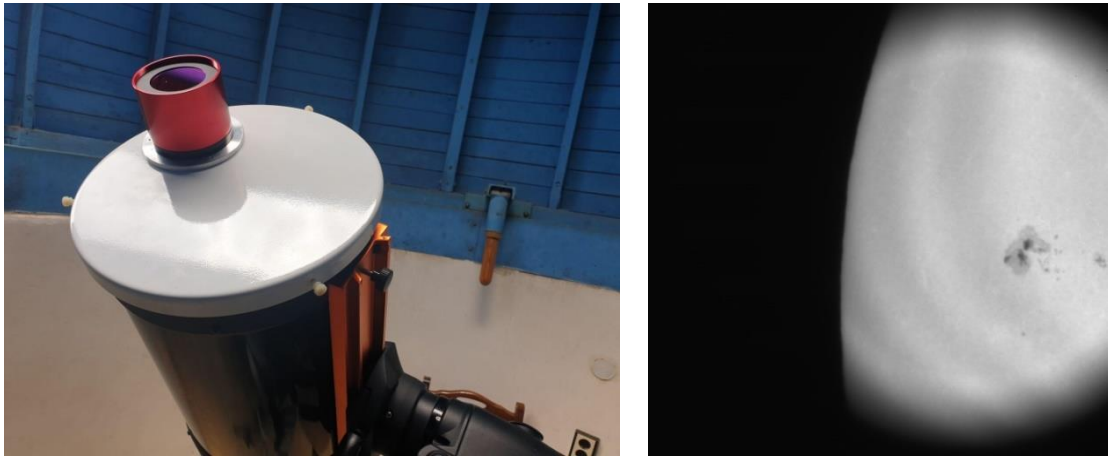

Fig. 1. Instrumentation and test image from AO-Belogradchik on 11-07-2025

The second station, Astronomical Center-Shumen University, is located on a plateau (at about 500 m elevation). The instrumentation used (Fig. 2, left) is as follows:

- Telescope: Coronado ST 40/400 PST Personal Solar Telescope OTA, D = 40 mm, focal ratio: f/10
- Mount/guiding: azimuthal, 'HelioFind' automatic Sun-guiding system
- Camera: NexImage 5 CCD camera
- Filter: integrated H-alpha

The test image shows a faint but clearly visible interference pattern, see Fig. 2 (right), with numerous sunspot groups (https://solarmonitor.org/index.php?date=20250923). The image was taken after a three-fold extension of the focal length to 1200 mm using a projection eyepiece.

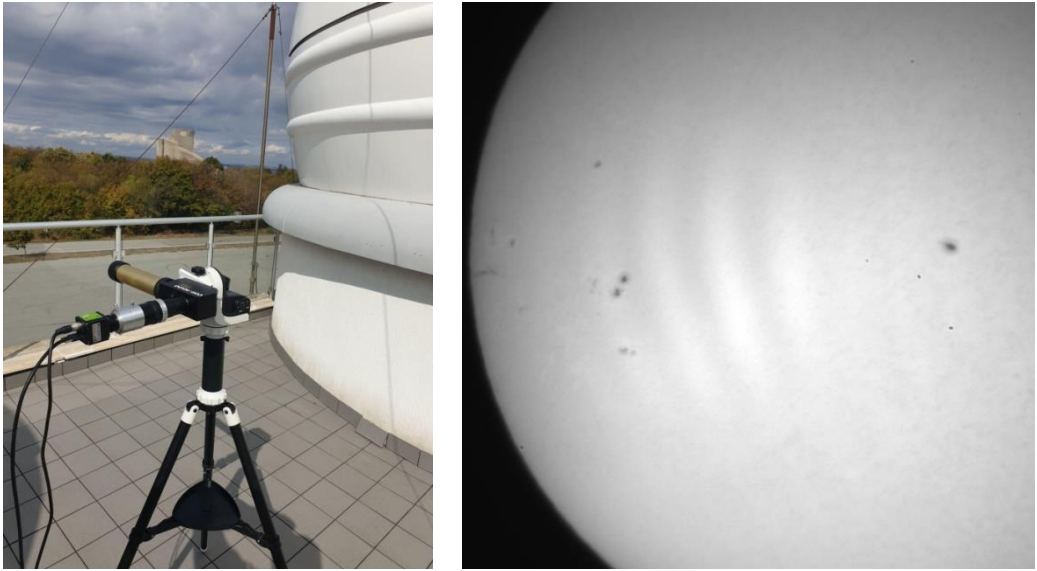

Fig. 2. Instrumentation at AO-Shumen and test image on 23-09-2025

The instrumentation to be used at the third station of the network, St. George RSP – Observatory (Fig.3, left), are described below:

- Guiding telescope: Refractor Celestron Omni XLT 120, D = 120 mm, f = 1000 mm, focal ratio: f/8.3
- Mount/guiding: 10Micron
- Camera: Atik Horizon Color CMOS Panasonic MN34230, chip size: 17.6 x 13.3 mm (22 mm diagonal), pixel size: 3.8 µm, pixel area: 4644 x 3506
- Filter: various metallized filters, glass and special film, i.e., AstroSolar Photo Film OD 5 from Baader

The performed tests show full-disk images, see an example in Fig. 3 (right) on 04-06-2025 (north is in the upper left side of the image). Several sunspot groups are visible at the western limb. According to the solar database https://solarmonitor.org/index.php?date=20250604 these are ARs 14100 (upper) and 14099 (lower).

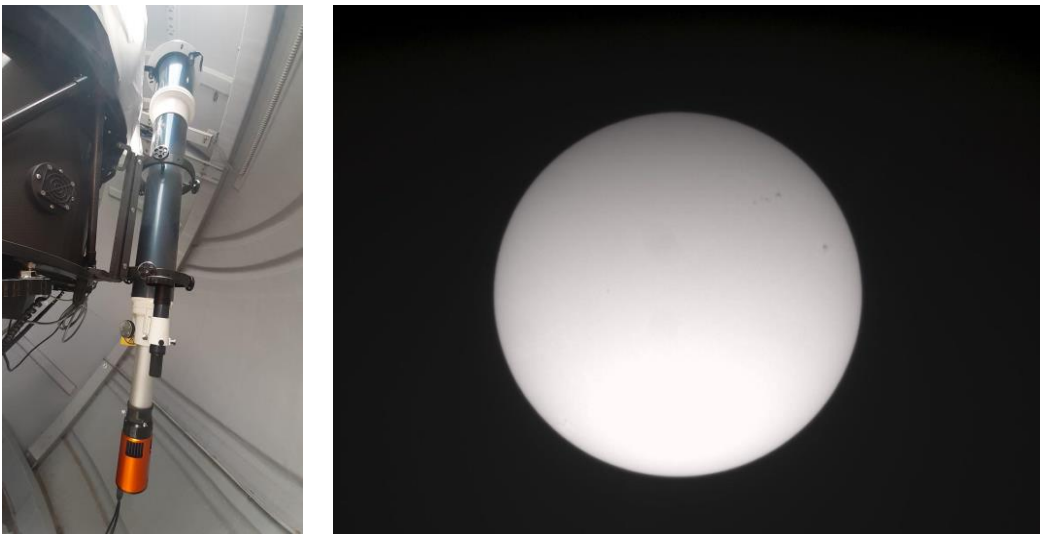

Fig. 3. Instrumentation and test image from St.George school and preschool observatory on 04-06-2025

**Outlook**

The future development aiming for an operational network includes:

- improvement of the optical systems and filters, incl. purchasing of new telescopes, rejection/blocking (white-light) and H-alpha filters if needed at either location;
- development of data transfer between the observatories, data storage, automatic pipelines for image processing;
- development and release of a public online database for solar monitoring in Bulgaria.

*Acknowledgments:*

*Selected results from this work have been obtained under the bilateral project "Joint Observations and Investigations of Solar Chromospheric and Coronal Activity" (https://astro.bas.bg/project-sun/) and supported by the Bulgarian National Science Foundation project No. KP-06-Austria/5 (14-08-2023) and Austria's Agency for Education and Internationalisation (OeAD) project No. BG 04/2023. DM acknowledges project RACIO supported by the Ministry of Education and Science of Bulgaria (Bulgarian National Roadmap for Research Infrastructure).*